\documentstyle [12pt]{article}

\newcommand{\be}{\begin{eqnarray}}
\newcommand{\ee}{\end{eqnarray}}
\newcommand{\BE}{\begin{equation}}
\newcommand{\EE}{\end{equation}}
\newcommand{\bt}{\begin{table}}
\newcommand{\et}{\end{table}}
\begin{document}
\baselineskip 0.33in

\begin{center}
\ \\
\ \\
{\Large{\bf Is Schr\"{o}dinger's Conjecture for the Hydrogen
Atom Coherent States Attainable?}}
\  \\
\vspace{0.5cm}
\  \\
{Ivaylo Zlatev$^{1}$, Wei-Min Zhang$^{2}$ and Da Hsuan Feng$^{1,3}$}\\
\vspace{0.4cm}
{\small{\em $^{1}$Department of Physics, Drexel University,
Philadelphia, PA 19104--9984 USA}} \\
{\small{\em $^{2}$Institute of Physics, Academia Sinica, Taipei 11529,
	Taiwan, ROC}} \\
{\small{\em $^{3}$Physics Division, Oak Ridge National Laboratory,
Oak Ridge, TN 37831 USA}} \\
\ \\
\ \\
\today
\end{center}
%++++++++++++++++++++++++++++++++++++++++++++++++++++++++++++++++++++++
\begin{abstract}
We construct the most general SO(4,2) hydrogen atom coherent states
which are the counterpart of Schr\"{o}dinger's harmonic oscillator coherent
states. We show that these states cannot be localized and cannot follow
the classical orbits. Thus, Schr\"{o}dinger's conjecture for the
hydrogen atom coherent states is unattainable.
\end{abstract}
%+++++++++++++++++++++++++++++++++++++++++++++++++++++++++++++++
\newpage

The classical limit of any quantum system is an important open question.
A natural way to realize the quantum to classical transition, first
proposed by Schr\"{o}dinger in 1926\cite{schrodinger}, is via
coherent states \cite{zha90}. These states, at least for the harmonic
oscillator (HO), are well-defined localized wavepackets that move along
the corresponding classical orbits. In the last paragraph of
his paper, Schr\"{o}dinger postulated that perhaps states with similar
properties may be constructed for the other simplest yet
fundamental quantum system, the hydrogen atom, and that the
wavepackets for this system would travel along Kepler elliptic
orbits. He pointed out that ``the technical difficulties in the
calculation (were) greater'' than in the HO case.
However, sixty-seven years later, such states {\it have not} yet been
constructed even though there were numerous attempts.
The main conclusion of this paper is to point out that
in principle such states as envisioned by Schr\"{o}dinger
{\it do not exist}.

The HO coherent states Schr\"{o}dinger constructed are linear
superpositions of all eigenstates of the HO which
form the irreducible representation space (Fock space) of the
Heisenberg-Weyl group $H_4$ (the dynamical group of the HO)
\cite{zha90}. All hydrogen atom bound eigenstates
(including all the degenerate states) form an irreducible
representation space of the SO(4,2) dynamical group \cite{Barut71}.
Thus, to find the analogous hydrogen atom coherent states that are linear
superpositions of all its eigenstates, it is natural to
use the group theoretical approach to construct the SO(4,2)
coherent states for the hydrogen atom.

A set of coherent states were previously constructed from the SO(4,2)
group and it was claimed that they follow the Kepler
motion \cite{mostowski}.  The validity of this statement
was later questioned \cite{stroud}. There were of course other
attempts to construct the coherent states of the hydrogen
atom by the use of wavepackets which did not utilize the above
irreducible representation space of SO(4,2) \cite{brown,stroud91,mike}.
In this paper we will show that the SO(4,2) coherent states of the
hydrogen atom can neither satisfy the general minimum uncertainty
relation, nor follow the Kepler orbits. Since these are the most
general coherent states for the system, we are compelled to conclude that
the hydrogen atom counterpart of Schr\"{o}dinger's HO coherent states
cannot possess the same physical features as the HO states.

The SO(4,2) group is a 3-rank, 15-dimensional non-compact semisimple
Lie group.  Its generators in the hydrogen atom representation are
\cite {Barut71}:
\be
	& & {\bf L} = {\bf r} \times {\bf p}, ~~ {\bf A} = \frac{1}{2}
		{\bf r}p^{2} - {\bf p} ( {\bf r} \cdot {\bf p} )
		- \frac{1}{2} {\bf r},  \nonumber       \\
	& & {\bf M} = \frac{1}{2} {\bf r} p^{2} - {\bf p} ( {\bf r} \cdot
		{\bf p} ) + \frac{1}{2} {\bf r},
		 ~~ {\bf \Gamma} = r{\bf  p},	         \\
 	& & S = \frac{1}{2} ( r p^{2} - r ), ~~ T = {\bf r} \cdot {\bf p} 		- i,
	~~ \Gamma_0 = \frac{1}{2} ( r p^{2} + r ). \nonumber
\ee
The irreducible representation space for the hydrogen atom
is a discrete lower-bound space with basis $|nlm \rangle$
which are the eigenstates of the hydrogen atom.  The lowest
state is $ |\psi_0 \rangle \equiv | n=1, l=0, m=0 \rangle $.
The maximum subgroup of SO(4,2) which leaves
the ground state invariant is $SO(4) \otimes SO(2)$.  Its generators are
$ L_i$, $A_i$ and $\Gamma_0$.  According to the general
algorithm \cite{zha90}, the $SO(4,2)$ coherent states
for the hydrogen atom are then defined as:
\be
 	|SO(4,2), CS \rangle &\equiv& |\Omega \rangle
		=  \Omega | \psi_0 \rangle \nonumber \\
	  &=& \exp i\left({\bf \alpha} \cdot {\bf M} + {\bf \beta}
		\cdot {\bf \Gamma} + \alpha_4 S + \beta_4 T \right)
		| \psi_0 \rangle,
\ee
where $\{ \alpha_i, \beta_i ; (i =1,2,3,4) \}$ are real parameters.

These coherent states for the hydrogen atom have useful
algebraic and topological properties. Just as the HO
coherent states are isomorphic to the coset space $H_4/U(1)
\otimes U(1)$, these states are isomorphic to the coset space
$SO(4,2)/SO(4) \otimes SO(2)$, a 4-dimensional complex manifold
which provides the geometrical basis for the link with the classical
limit of the hydrogen atom. Their cardinal algebraic property is
the resolution of identity:
\BE
	\int d\mu | \Omega \rangle \langle \Omega | = I
\EE
where $d\mu$ is the normalized $SO(4,2)$-invariant measure on
$SO(4,2)/SO(4) \otimes SO(2)$.  This property offers a
connection between the quantum observables and the classical
observables \cite{Klauder}. We mentioned that there were other
constructions of wavepackets for
the hydrogen atom \cite{brown,stroud91,mike}.
These wavepackets
do not satisfy the above resolution of identity \cite{Klauder}.

One can physically appreciate the coherent states we have constructed
for the hydrogen atom by comparing them to the Schr\"{o}dinger's  HO
coherent states.  It is now well known that the Schr\"{o}dinger's HO
coherent states can be written as \cite{foot1}:
\BE
	| z \rangle = \exp (za^{\dagger} - z^* a)| \psi_0 \rangle
		= e^{-|z|/2} \sum_{n=0}^{\infty}\frac{z^n}{\sqrt{n!}}
		| n \rangle
\EE
where $|n \rangle$ are the HO eigenstates.  These coherent
states are wavepackets and are well localized at the point
$z$ in the HO phase space.  This is a direct
result of the fact that $\frac{z^n}{\sqrt{n!}}$ has its maximum
at $|z|^2=n$, as Schr\"{o}dinger pointed out in his paper.
Hence, in the expansion, the coherent states will be dominated
by a small number of eigenstates in the neighborhood of
this $|z|^2$. Here  we must point out
that this is true not only for very large $|z|$ as Schr\"{o}dinger
had stated in his paper, but for any value of $|z|$ as well.
The rigorous statement of the above well-localized wavepackets
is that they will satisfy the minimum uncertainty relation:
\BE
	\Delta q \Delta p = \hbar/2.
\EE

Schr\"{o}dinger stated in his paper \cite{schrodinger} that ``{\it das man
auf ganz \"{a}hnliche Weise auch die Wellengruppen konstruieren
kann, welche auf hoch-quantingen Keplerellipsen umlaufen und das
undulationsmechanische Bild des Wasserstorffelektrons
sind.}'' \cite{trans}. The word  ``{\it hoch-quantingen}''
means ``high quantum number'' \cite{thank}.  The same sort of
statement about the HO case was made earlier by Schr\"{o}dinger in his
paper, and we now see that
in fact it is not necessary for the HO coherent states. Hence, we
should not expect that the
equivalent hydrogen atom coherent states need to be constructed
from eigenstates with high quantum numbers.

To compare the above picture of the HO coherent states
to that of hydrogen atom coherent states, it is convenient to use the boson
representation of the SO(4,2) algebra \cite{Barut71}:
\be
	&& L_k = \frac{1}{2}\epsilon_{ijk}({\bf a}^{\dagger}\sigma_k{\bf a}
                    + {\bf b}^{\dagger}\sigma_k{\bf b} ),
	 ~~~A_i = -\frac{1}{2}({\bf a}^{\dagger}\sigma_i{\bf a}
                    - {\bf b}^{\dagger}\sigma_i{\bf b}), \nonumber\\
	&& M_i = -\frac{1}{2}({\bf a}^{\dagger}\sigma_iC{\bf b}^{\dagger}
                    - {\bf a}C\sigma_i{\bf b}),
	 ~~~\Gamma_i = -\frac{i}{2}({\bf a}^{\dagger}\sigma_iC{\bf b}^{\dagger}
                    + {\bf a}C\sigma_i{\bf b}), \\
	&& S = \frac{1}{2}({\bf a}^{\dagger}C{\bf b}^{\dagger}
                    + {\bf a}C{\bf b}),
	~~~ T = -\frac{i}{2}({\bf a}^{\dagger}C{\bf b}^{\dagger}
                    - {\bf a}C{\bf b}),
	~~~ \Gamma_0 = \frac{1}{2}({\bf a}^{\dagger}{\bf a}
                    + {\bf b}^{\dagger}{\bf b} +2)  \nonumber,
\ee
where $\sigma_i$ are the Pauli spin matrices, $C=i\sigma_2$ and
${\bf a}=(a_1,a_2)$, ${\bf b}=(b_1,b_2)$ are the two harmonics
oscillator creation and annihilation operators.  The hydrogen
atom ground state is the boson vacuum state, $ a_i | \psi_0
\rangle = b_i | \psi_0 \rangle = 0 ~~, i =1,2$, and its
coherent states, Eq.(2), can then be reexpressed as
\BE
 	|\Omega \rangle	= \exp \sum_{i,j=1}^{3} \left(z_{ij} a^{\dagger}_i
	          b^{\dagger}_j - z_{ij}^* a_ib_j \right)
		| \psi_0 \rangle,  \label{bcs}
\EE
where the relations between $z_{ij}$ and $\{\alpha_i, \beta_i \}$ can be
found by comparing Eq.(2) and Eq.(7).
The coherent states can further be rewritten as
\be
	&& |\Omega \rangle
	     =  {\cal N}^{1/2} \sum_{n=1}^{\infty} \sum_{k_1 k_2 k_3 k_4}
		C_{1234}\frac{(z_{11}^{k_1} z_{12}^{k_2} z_{21}^{k_3} z_{22}^
		{k_4})}{(n-1)!} | n \rangle          \nonumber  \\
	&& ~~~~~=  {\cal N}^{1/2} \sum_{n=1}^{\infty} \sum_{lm}
		C_{lm}^{z} | nlm \rangle  \label{csep}
\ee
where $\cal N$ is the normalized constant, $k_1 + k_2 + k_3 + k_4 =
 n-1$,~~~~~~~~
and
$ | n \rangle = (a^{\dagger}_1)^{k_1+k_2}
(a^{\dagger}_2)^{k_3+k_4}(b^{\dagger}_1)^{k_1+k_3} (b^{\dagger}_2)^{k_2+k_4}
| \psi_0 \rangle$ which are the eigenstates of the hydrogen atom with the
principal quantum number $n$
\BE
	\Gamma_0 | n \rangle = n | n \rangle.
\EE

Using Schr\"{o}dinger's argument for the HO coherent
states, we can see that the SO(4,2) coherent states
for hydrogen atom are not well-localized wavepackets.
This is because from Eq.(\ref{csep}) it is easy to see that
for given $z_{ij}$, there is no unique peak as in the HO case.
On the contrary, there is a large number of peaks which
will {\it de facto} delocalize the wavepackets.  Hence
it is not expected that the SO(4,2) coherent states for
the hydrogen atom can satisfy the minimum uncertainty relation.

Next we will show rigorously that the SO(4,2) coherent
states cannot satisfy the general minimum uncertainty:
\BE
	(\Delta r)^2(\Delta p_r)^2 = \frac{1}{4}\langle \Omega|
		[r-\langle r\rangle,p_r-\langle p_r\rangle]_+|
		\Omega \rangle^2 + \frac{\hbar^2}{4}, \label{gmuc}
\EE
of which the Heisenberg minimum uncertainty relation is a special
case. The proof is as follows. Eq.(\ref{gmuc}) is true only if
\cite{shankar}:
\BE
	(r-\langle r\rangle)|\Omega\rangle = c (p_r - \langle p_r\rangle)
		|\Omega\rangle,
\EE
which is the same as:
\BE
	(r^2-r\langle r\rangle)|\Omega\rangle = c({\bf r}\cdot{\bf p}
		- r\langle p_r\rangle)|\Omega\rangle.	\label{uc2}
\EE
However, we find that
Eq.(\ref{uc2}) cannot be satisfied by the $SO(4,2)$ coherent
states. Using the matrix representation approach in the coherent
state calculation \cite{zha90}, Eq.(\ref{uc2}) can be reduced to
\BE
	A | 0 \rangle + B | 2 \rangle + C | 4 \rangle = 0
\EE
where $ |0 \rangle \equiv | \psi_0 \rangle, | 2 \rangle \equiv
a_i^{\dagger}b_j^{\dagger} | \psi_0 \rangle$, and $ | 4 \rangle
\equiv a_i^{\dagger}b_j^{\dagger} a_k^{\dagger}b_l^{\dagger}
| \psi_0 \rangle$, and the coefficients $A, B, C$ are functions
of $z_{ij}$.  It can explicitly be shown that the coefficients
$A, B$ and $C$ vanish only if $C =\frac{ \langle r^2\rangle}
{\langle{\bf r}\cdot{\bf p}\rangle}$, and
\BE
	2Y_1X +iX \sigma_2X -iY_1\sigma_2Y_2 = 0,~~~~~
		X \sigma_2 X + Y_1 \sigma_2 Y_2 =0, \label{condi}
\EE
where
\BE
	X = \frac{Z}{\sqrt{Z^{\dagger}Z}}sinh\sqrt{Z^{\dagger}Z},
\EE
$Z$ is a $2 \times 2$ matrix with elements $z_{ij}$, the parameters
in the coherent states of Eq.(\ref{bcs}), and
\BE
	Y_1=\sqrt{I + XX^{\dagger}}, ~~~ Y_2=\sqrt{I + X^{\dagger}X}
\EE

Now using Eq.(\ref{condi}), we find that $\Delta r =0$, and
 $\Delta p_r$ is infinite, which seems to be consistent
with the uncertainty relation.  However, in this case,
the coherent state becomes an eigenstate of the $r$ operator
with zero eigenvalue and therefore contains no physical
information.  In fact, we find that Eq.(\ref{condi}) leads to
\BE
	-4Im^2(x_{12}) -1 = (Re(x_{11}) + Re(x_{22})^2 +
		Im(x_{11}) - Im(x_{22}))^2
\EE
a condition
which cannot be satisfied. In eq. (17), $Re(x_{ij})$ and $Im(x_{ij})$
mean the real and imaginary part of the matrix element $x_{ij}$ of the
matrix X. This indicates that Eq.(\ref{gmuc})
cannot be obeyed. Thus, the requirement that the SO(4,2)
coherent states satisfy the general minimum uncertainty
cannot be realized.

We conclude that the most general SO(4,2) coherent states
for the hydrogen atom cannot be well-localized wavepackets.
In fact, based on the recent work proposed by two of the authors
\cite{wz},  we can show that the coherent states that are the
superposition of all the eigenstates of the hydrogen atom
cannot give us the classical limit. The classical limit in
this framework is realized as follows: classical dynamics, if
it exists at all for a quantum system, is depicted by the
phase space structure of the expectation value of the hamiltonian
in the coherent states in which the quenching index $\Xi$,
which is inherently hidden in the coherent states, goes to
infinity.  In the SO(4,2) coherent states for the hydrogen
atom, $\Xi= \frac{1}{2}$. So, there is no classical limit
for this quantum system.  As a result, the coherent states
cannot move along the Kepler orbits. It would therefore appear
that the goal of Schrodinger is unattainable.

Finally, we should remark that the results we have
obtained here do not preclude that coherent states formed
by a  complete {\em subset} of the hydrogen eigenstates have a
classical limit. Such states are for example the SU(1,1)
coherent states, which are superpositions of the states with
$n \geq l +1$ where $l$ is the angular momentum quantum number
and should be a relatively large value since the quenching
index for SU(1,1) hydrogen atom coherent states is $\Xi= 2(l+1)$
\cite{wz}.  Quite recently,
there were some remarkable experimental developments
on single atoms and  radiation  {\cite {stroud,koch}} which
may imply that there will be some wavepackets which
will follow classical trajectories and have
both classical and quantum properties. Our work on
the SU(1,1) hydrogen atom coherent states is now in progress.
We speculate that these could be the SU(1,1) coherent states for the
hydrogen atom.

This work is supported by the U.S. National Science Foundation and
the National Science Council of the ROC. We thank B. Giraud for a useful
communication.

\newpage
%##########################################################
%##########################################################

\end{document}